\documentclass[manuscript]{rmaa}

\usepackage{natbib}
\title{Large \hi\ structures linked to southern O-type stars} 

\author{
  M. C. Mart\'{\i}n,\altaffilmark{1} 
  C. E. Cappa,\altaffilmark{1,2}  
  and G. A. Romero\altaffilmark{1,2}
  }
  
\altaffiltext{1}{ Instituto Argentino de Radioastronom\'{\i}a (CCT-La Plata, CONICET), C.C. 5, 1894, Villa Elisa, Bs. As., Argentina }

\altaffiltext{2}{Facultad de Ciencias Astron\'omicas y Geof\'{\i}sicas, Paseo del Bosque S/N, 1990, Universidad Nacional de La Plata, Argentina}

\shortauthor{Mart\'{\i}n, Cappa \& Romero}
\shorttitle{\hi\ structures linked to O-type stars}

\fulladdresses{
\item Instituto Argentino de Radiostronom\'{\i}a, Parque Pereyra Iraola Km 40,500,
  CC 5, Villa Elisa, Bs. As., Argentina}

\listofauthors{M. C. Mart\'{\i}n, C. E. Cappa \& G. A. Romero}
\indexauthor{Mart\'{\i}n, M. C.}
\indexauthor{Cappa, C. E.}
\indexauthor{Romero, G. A.}

\abstract{In our search for interstellar bubbles around massive stars we analyze 
the environs of the O-type stars HD\,38666, HD\,124979, HD\,163758, and
HD\,171589. The location of the stars, which are placed far from the 
galactic plane, favors the formation of large wind bubbles.
We investigate the distribution of the neutral and ionized gas based on 
\hi, CO, and radio continuum data, and that of the interstellar dust based 
on far infrared IRIS images.
Here we report the discovery of neutral gas 
cavities and slowly expanding shells associated with the four massive stars. IR and optical counterparts were also detected for some of the stars.
We discuss the probability that the features have originated in the action of the stellar winds on the surrounding gas.}

\resumen{Analizamos la distribuci\'on del material interestelar en la vecindad de las estrellas de tipo O: HD\,38666, HD\,124979, HD\,163758 y HD\,171589, a fin de investigar la existencia de burbujas interestelares asociadas a las estrellas. La localizaci\'on de las estrellas lejos del plano gal\'actico, favorece la formaci\'on de burbujas de gran extensi\'on. En base a datos del \hi, continuo de radio y CO, e im\'agenes en el infrarrojo lejano que revelan la emisi\'on del polvo interestelar, investigamos la distribuci\'on del gas neutro e ionizado en estas regiones. Reportamos en este trabajo la detecci\'on de cavidades y envolturas en expansi\'on en hidr\'ogeno neutro, asociadas a las cuatro estrellas masivas. Se discute la posibilidad de que estas estructuras hayan sido originadas por la acci\'on de los vientos estelares sobre el gas circundante.}

\addkeyword{ISM: bubbles} 
\addkeyword{stars: individual (HD\,38666, HD\,124979, 
HD\,163758, HD\,171589)}

\begin{document}
%
\def\kms{km s$^{-1}$}
\def\etal{et al.}
\def\hi{H{\sc i}}
\def\hii{H{\sc ii}}
\def\deg{$^\circ$}
\def\msun{M$_\odot$}
\def\lsun{L$_\odot$}
\def\mjyb{mJy beam$^{-1}$}
\def\jyb{Jy beam$^{-1}$}
\def\mdot{M$_\odot$ yr$^{-1}$}
\def\ldot{L$_\odot$}
\def\lst{L$_{*}$}
\def\cmtres{cm$^{-3}$}
\def\cmdos{cm$^{-2}$}
\def\ojo{\fbox{\bf !`$\odot$j$\odot$!}}
\def\ap{$\approx$}
\maketitle

\section{Introduction}
\label{sec:intro}

The strong stellar winds of massive stars interact with their environs 
creating {\it interstellar bubbles} that are detected in a large range 
of wavelengths. UV photons with energies h$\nu \geq$ 13.6 
eV ionize the inner part of the expanding bubbles, which are 
generally detected both 
in the optical \citep{Loz82} and in the radio continuum ranges as 
shell-like thermal sources \citep{GL95}. Should the ionizing 
front be trapped within the expanding envelope, interstellar bubbles have 
an outer neutral region that can be identified both in the \hi\ 21--cm  line 
emission and in molecular lines \citep{CR05} 

Many interstellar bubbles are detected neither in the radio continuum
nor in optical lines (e.g. \citealp{CB98,Ar99,CiA04}). A low ambient density 
may be responsible for the undetected radio continuum or optical emission,
as was suggested for bubbles in the LMC by \citet{Naz02}. As a consequence, 
the analysis of the \hi\ emission distribution is an important tool to 
investigate the presence of such structures and allows discrimination in 
distance based on galactic rotation models. 

Data at radio and infrared wavelengths offer an opportunity to investigate the
characteristics and distribution of the ionized and neutral gas associated 
with these bubbles, as well as those of the interstellar dust. 

A low-velocity massive  star with constant stellar wind parameters located in 
an homogeneous interstellar medium with 
uniform density creates a spherical stellar wind bubble.
Aspherical bubbles appear as the result of relaxing some of these
hypothesis.
The combination of a powerful stellar wind and a high space velocity
makes this scenario changes drastically. \citet{We77} pointed out 
that a massive star moving supersonically with respect to the surrounding 
gas originates an aspherical
stellar bubble, elongated in the direction of the movement of the star.
A wind bow-shock can develop ahead of the star \citep{Wi96,Rag97}.
As \citet{vBMc88} have shown, these shocks can be identified in 
the far IR emission.

Here we present a large scale study of the interstellar medium around 
four southern 
O-type stars: HD\,38666, HD\,124979, HD\,163758, and HD\,171589, the first two of
them classified as runaway stars. Our aim 
is to investigate the action of the UV photon flux and the stellar 
winds of these stars on their surroundings. The selected stars
are located at $|${\it b}$| >$ 3\deg, allowing for the formation of
relatively large structures which are easily identified using low angular
resolution data.

In the following sections, we analyze the interstellar medium in the 
environs of the stars looking for cavities and shells in the neutral gas
that have originated in the action of the massive stars
on their surroundings. These studies are important to improve theoretical 
models on the interplay between massive stars and the surrounding gas.

\begin{table}[!t]\centering
  \small
  \newcommand{\DS}{\hspace{6\tabcolsep}} 
  \begin{changemargin}{-2cm}{-2cm}
   \caption{Main parameters of the O-Type stars} \label{tab:ion_ab}
    \setlength{\tabnotewidth}{0.95\linewidth}
    \setlength{\tabcolsep}{0.5\tabcolsep} \tablecols{10}
    \begin{tabular}{l @{\DS} cccc l cccc}
    \toprule
(1) & (2) & (3) & (4) & (5) & (6) & (7) & (8) & (9) & (10) \\ 
    \midrule
Name & {\it (l,b)} & Spectral & $d$ & $V$ & $(B-V)$ & $d_{spc}^{\mathrm{e}}$  &  $\mu_l \ cos b^{\mathrm{f}}$ &  $\mu_b^{\mathrm{f}}$ & z\\
     &       & Classification \\
 & (\deg) & & (kpc) &  (mag) & (mag) & (kpc) & (mas yr$^{-1}$) &  
(mas yr$^{-1}$) & (pc) \\
      \midrule  
 HD 38666 &  237.29,--27.10  &  O9.5 V$^{\mathrm{a}}$  & 0.53$^{\mathrm{a}}$ & 5.17$^{\mathrm{a}}$ & --0.27$^{\mathrm{a}}$ & 0.8 &  +23.48$\pm$1.1 &  --3.52$\pm$1.1 & --255\\
HD 124979 &  316.40,+9.08 &  O8 III ((f))$^{\mathrm{b}}$ &  & 8.52$^{\mathrm{a}}$ & 1.05$^{\mathrm{a}}$ & 3.7 & --8.8$\pm$1.1 &  +8.2$\pm$1.1 & +590\\
HD 163758 & 355.36,--6.10 & O6.5 Ia (f)$^{\mathrm{a}}$ & & 7.32$^{\mathrm{a}}$ & 0.03$^{\mathrm{a}}$ & 3.5 & --2.5$\pm$1.5 & --5.2$\pm$1.5 & --374\\
HD 171589 & 18.65,--3.09 & O7 II (f)$^{\mathrm{a}}$ &  1.15$^{\mathrm{c}}$-1.5$^{\mathrm{d}}$ & 8.29$^{\mathrm{a}}$ & 0.29$^{\mathrm{a}}$ & 3.0 &  +6.0$\pm$1.5 &  --4.1$\pm$1.5 & +160\\
    \bottomrule
    \tabnotetext{a}{\small GOS catalogue (\citet{Maiz-Apellaniz04})}
\tabnotetext{b}{See text}
\tabnotetext{c}{\small \citet{Ko85}}
\tabnotetext{d}{\small \citet{Gar82}}
\tabnotetext{e}{\small Spectrophotometric distances derived in this paper. See text for details}
\tabnotetext{f}{\small Derived from Tycho-2 catalogue \citep{Hog00}.}
  \end{tabular}
  \end{changemargin}
\end{table}

\section{Data bases}
\label{sec:data}

The \hi\ 21--cm  line data analyzed in this paper were extracted from the
Leiden Argentine Bonn Survey of Galactic \hi\ \citep{KB05,Arnal2000} observed using the radiotelescope of the Instituto
Argentino de Radioastronom\'{\i}a. The 21--cm data span the velocity range from
--250 to +250 \kms\ and were obtained with a velocity resolution of 1.3 \kms. The angular resolution is 30\arcmin.

Radio continuum data at 4.85 GHz from the Parkes-MIT-NRAO (PMN)
Southern Radio Survey \citep{C93} are available only for HD\,124979 and HD\,171589, while data at 408 MHz \citep{Has82} and at 35 MHz \citep{Dwarakanath90} are available for the four target stars. The angular resolutions
are 5$\arcmin$, 0\fdg 85, and 26$\arcmin$ x 42$\arcmin$/cos($\delta$-14\deg), for the surveys at 4.85 GHz, 408 MHz, and 35 MHz, respectively.

The analysis of the molecular gas distribution for the regions
of HD\,163758 and HD\,171589 was based on the CO survey
by \citet{Dame01}, which has an angular resolution of 8\farcm 8.
The velocity coverage and the velocity resolution are --160 to +160 \kms\ 
and 1.3 \kms, respectively. The rms noise level is 0.2 K.

The dust distribution in the region was analyzed  using the 
IRIS\footnote{http://www.ias.fr/IRIS} database. These images are a new 
generation of IRAS images that benefit from a better zodiacal light 
subtraction, calibration, and a better destriping. IRIS images have an
angular resolution of around 4\arcmin\ \citep{ML05}. 

The optical images were obtained from the Full-Sky H-Alpha survey (H-Alpha 
Composite, \citet{Fi03})

\section{Target stars}

Table 1 summarizes the stellar parameters relevant to this study: the 
name of the stars and their galactic coordinates are listed in the first 
two columns. The spectral types, indicated in the third column,  were 
obtained from \citet{Maiz-Apellaniz04}(GOS catalogue). 
Distance estimates taken from the literature are given in column 4.
Columns 5 and 6 list the visual magnitudes $V$ and color indices $(B-V)$, respectively.
We used these values together with absolute magnitudes from \citet{Va96} and 
intrinsic color indices from \citet{Wegner94}
 to derive the
spectrophotometric distance listed in column 7. The components of 
the proper motion in the galactic coordinate system listed in columns 8 and 
9 were calculated from the Tycho-2 catalogue \citep{Hog00}.
Finally, z-distances to the galactic plane are shown in col. 10. They were derived by adopting the distances
listed in col. 7.

Based on its large proper motion, HD\,38666 ($\mu$ Col) was identified
as a runaway star belonging to Ori\,OB1 \citep{Blaauw61}.

HD 124979 has been classified as an O8 ((f)) star by \citet{MaBi76}.
Following the Of typification by \citet{Wa71}, we have
adopted a luminosity class III. \citet{Mason98} analyzed the
binarity status, and concluded that it maybe a spectroscopic
binary, tagging it as ``SB1?''.
The measured proper motion and radial velocity ($-$68 km s$^{-1}$) allow
to classify this star as a runaway.

No information about the interstellar medium in the environs of the target
stars was found in the literature.

\section{\hi\ emission towards the selected stars}

Figure 1 ($\it{a}$-$\it{d}$) displays the \hi\ 21--cm  line profiles towards
the O-type stars. They were obtained by averaging the \hi\ emission within
an area of  about  3\deg $\times$3\deg\ centered at the position of the
massive stars. Our aim is to analyze the main characteristics of the
neutral gas emission towards the selected stars and its relation to the
galactic spiral structure.

The \hi\ emission profile in the region of HD\,38666 (Fig. 1a) reveals
a weak \hi\ gas component centered at $\rm{v} \approx$ 0 \kms\ (all
velocities in this paper are referred to the LSR). Because of
the high galactic latitude of the star, this material corresponds to
local gas. Bearing in mind the distances to HD\,38666 listed in Table 1,
we expect that gas related to this star has low positive velocities.

\begin{figure*}
   \begin{center}
   \includegraphics[width=0.9\textwidth]{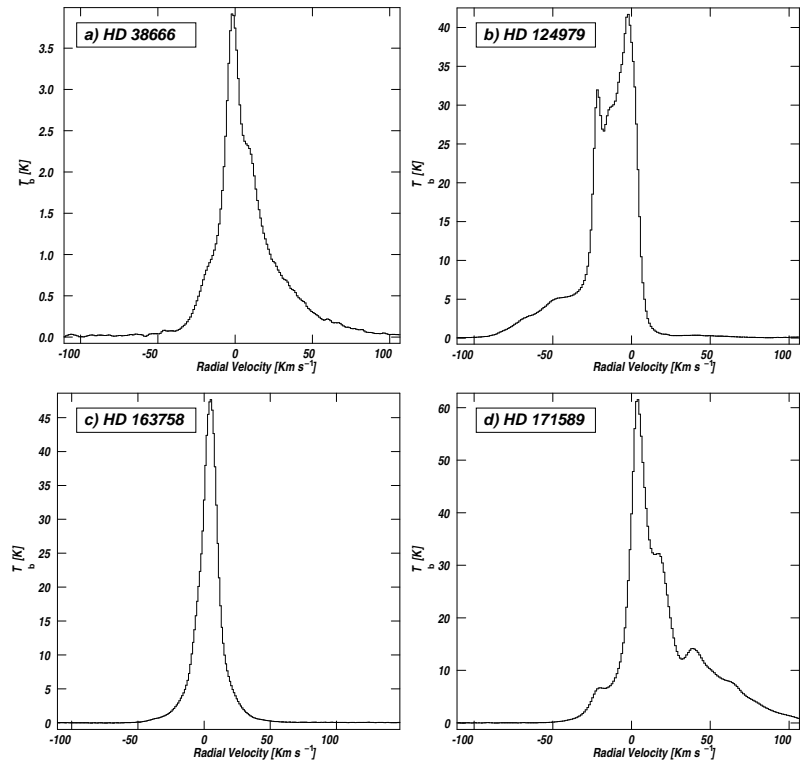}
   \caption{\hi\ profiles obtained by averaging the \hi\ emission in
regions of 3\deg $\times$3\deg\ around the O-type stars.
{\it (a)} HD\,38666;  {\it (b)} HD\,124979;
{\it (c)} HD\,163758; and {\it (d)} HD\,171589. The intensity scale is given
in brightness temperature. Velocities are referred to the LSR.}
   \end{center}
   \end{figure*}

The \hi\ emission profile in the region of HD\,124979 (Fig. 1b) reveals
three components centered at $\rm{v} \approx$ --2,  $\rm{v} \approx$ --20,
and $\rm{v} \approx$ --50 \kms. The former gas component is associated with
the local region. The circular galactic rotation model by
\citet{BB93} predicts near and far kinematical distances
$d \approx$ 1.4 and 11 kpc, respectively, for the component at --20 \kms,
while for the component at --50 \kms, $d \approx$ 3.7 and 8.5 kpc,
respectively. The near kinematical distances  are compatible with the
locations of the Sagittarius-Carina and Scutum-Crux spiral arms
\citep{Ru03}. According to the analytical fit by \citeauthor{BB93}, gas in
the environs of this star, i.e. at 3.7 kpc should have velocities of
about --50 \kms.

In the line of sight to HD\,163758 (Fig. 1c), \hi\ emission was detected
from --40 to +40 km$^{-1}$. The profile shows a main peak of
neutral gas emission centered at $\rm{v} \approx$ +5 \kms, probably
related to the local spiral arm. The fact that positive velocities
are forbidden in the fourth galactic quadrant within the solar circle
(e.g. \citealp{BB93}) indicates that most of this material is local and
associated with the Gould's belt \citep{Olano82}. Gas linked to this
star at a distance of 3.5 kpc should have velocities of about --10 \kms.
Note however that kinematical distances have large uncertainties in
this line of sight.

The \hi\ emission towards HD\,171589 (Fig. 1d) was detected within the
velocity interval --40 to +100 \kms. The neutral gas emission shows major
peaks at $\rm{v} \approx$ 0, $\rm{v} \approx$ +12 \kms, and
$\rm{v} \approx$ +40 \kms. The lower velocity peak belongs to the local
spiral arm, while peaks at positive velocities correspond to near and far
kinematical distances $d \approx$ 1.3 and 15 kpc, and  $d \approx$ 3.4 and 12.5
kpc, respectively. These gas components are probably related to the
Sagittarius-Carina and Scutum-Crux spiral arms, respectively (Russeil 2003).
A relatively low emission component can also be identified at
$\approx$ --25 \kms. The analytical fit by \citeauthor{BB93} predicts that gas
with this velocity is located at $\approx$ 20 kpc, well beyond the solar
circle. Material related to the star, placed at $d \leq$ 3 kpc, should
appear with velocities lower than +33 \kms.

\section{\hi\ structures and their counterparts at other frequencies}

In order to look for \hi\ structures linked to the selected stars, we
analyzed the neutral hydrogen distribution in their environs. A series of
$\it{(l,b)}$ maps at maximum velocity resolution (1.3 \kms) showing the
\hi\ emission distribution at different velocities were constructed.
The \hi\ emission is shown in Fig. 2 for HD\,38666, Fig. 4 for HD\,124979, Fig. 6
for HD\,163758, and Fig. 8 for HD\,171589. Each individual image
corresponds to a velocity interval of 4 \kms. Although the whole velocity
range at which \hi\ emission is detected was analyzed, the figures
include only the images where \hi\ structures probably linked to the stars 
are detected. 

The identification of an \hi\ structure associated with a certain star 
requires a careful inspection of the \hi\ images, looking for shells and 
voids that might be related to the star.
Several conditions are necessary to associate an \hi\ structure with 
a certain star:
(a) the star is expected to be located inside the \hi\ void or close to 
its  inner borders; (b) the kinematical distance to the structure should 
coincide, whithin errors, with the spectrophotometric distance; and 
(c) bearing in mind that the velocity dispersion in the interstellar medium 
is about 6 \kms, an \hi\ feature should remain detectable for a 
larger velocity interval to be considered as a physical structure. 

We have found a number of \hi\ structures towards the target stars that 
fulfill these conditions and, consequently, can be related to the stars.
In the following sections we describe the \hi\ structures along with 
their counterparts at other wavelengths.

The main physical parameters of each structure are summarized in Table 2, which is described in \S\ 6.1.

   \begin{figure}
   \centering
    \includegraphics[width=.95\textwidth]{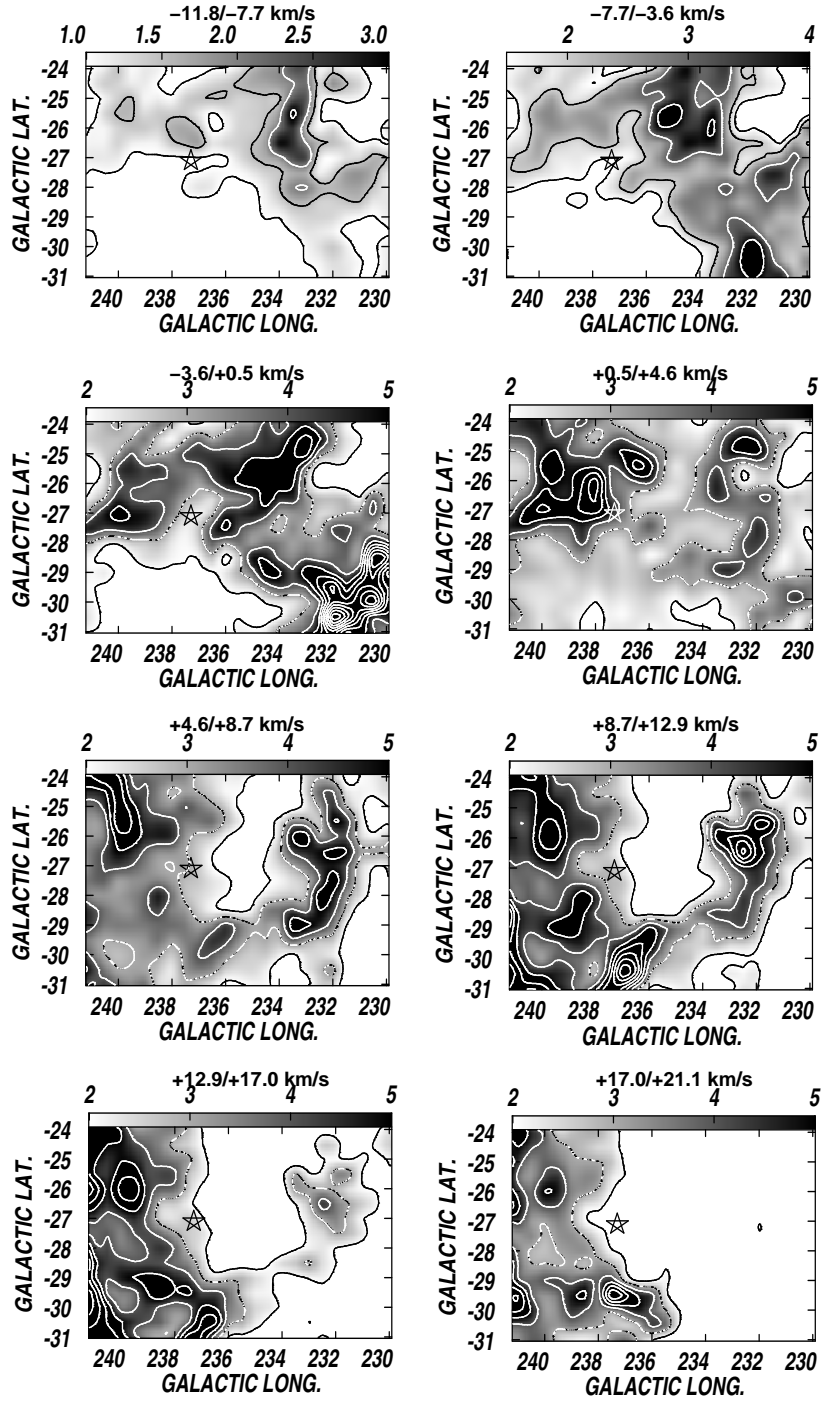}
   \caption{\hi\ emission distribution towards HD\,38666, integrated in 
steps of 4 \kms. The velocity range and the greyscale (in K) of each image are
indicated in its upper part. The contour lines are from the minimum 
value of the grayscale to 12 K, in steps of 1 K. HD\,38666 is indicated by the
star. }
    \end{figure}
        
 \begin{figure}
   \centering
          \includegraphics[width=.75\textwidth]{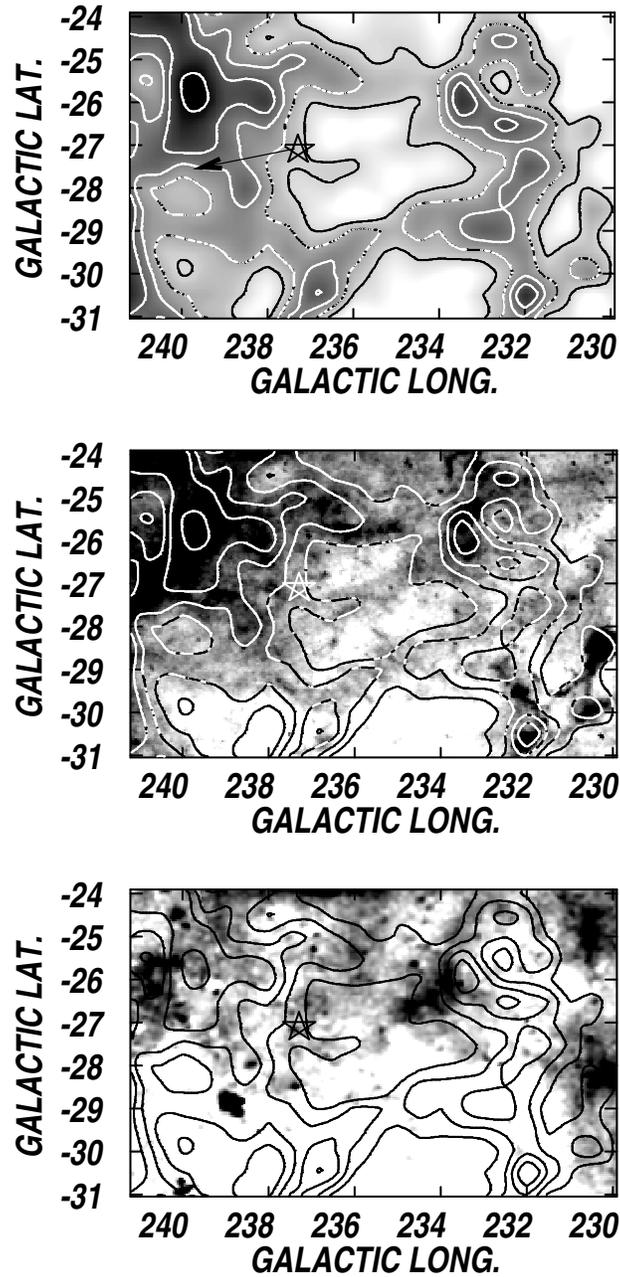}\\
          \includegraphics[width=.75\textwidth]{HD38666-IRIS-100.EPS}\\                  
          \includegraphics[width=.75\textwidth]{HD38666-HALPHA-HI.EPS}\\
          \caption{{\it Upper panel.} \hi\ emission distribution within the velocity 
interval --7.2 to +17.0 \kms\ showing the \hi\ structure 
related to HD\,38666. 
The grayscale is from 2 to 5 K, and the contour lines are 2.5, 3, 3.5, 4 and 5 K. HD\,38666 is indicated by a star. 
{\it Central panel.} Overlay of the IR emission at 100 $\mu$m (grayscale)
and the same \hi\ contours of the upper panel. The grayscale corresponds to 
1.8 to 2.8 MJy ster$^{-1}$).
{\it Bottom panel.} Overlay of the H-Alpha Composite image (grayscale) and 
the same \hi\ contours of the upper panel. The grayscale is from 4.3 to 6.5 R.
}
\end{figure}

\subsection{The ISM around HD\,38666}

The \hi\ emission distribution towards this star within the velocity 
interval from --11.8 to +21.1 \kms\ is displayed in Fig. 2. HD\,38666 is 
indicated by a star. The images show that the O-type star is projected 
onto the border of a low emission region identified within the velocity 
interval from --7.7 to +17.0 \kms, centered approximately at {\it (l,b)} = 
(235\fdg 5,--27\fdg 5). Identification of \hi\ structures is difficult 
for velocities higher than +17 \kms\ due to the lower general \hi\ emission 
in the region.

   \begin{figure}
   \centering
          \includegraphics[width=.95\textwidth]{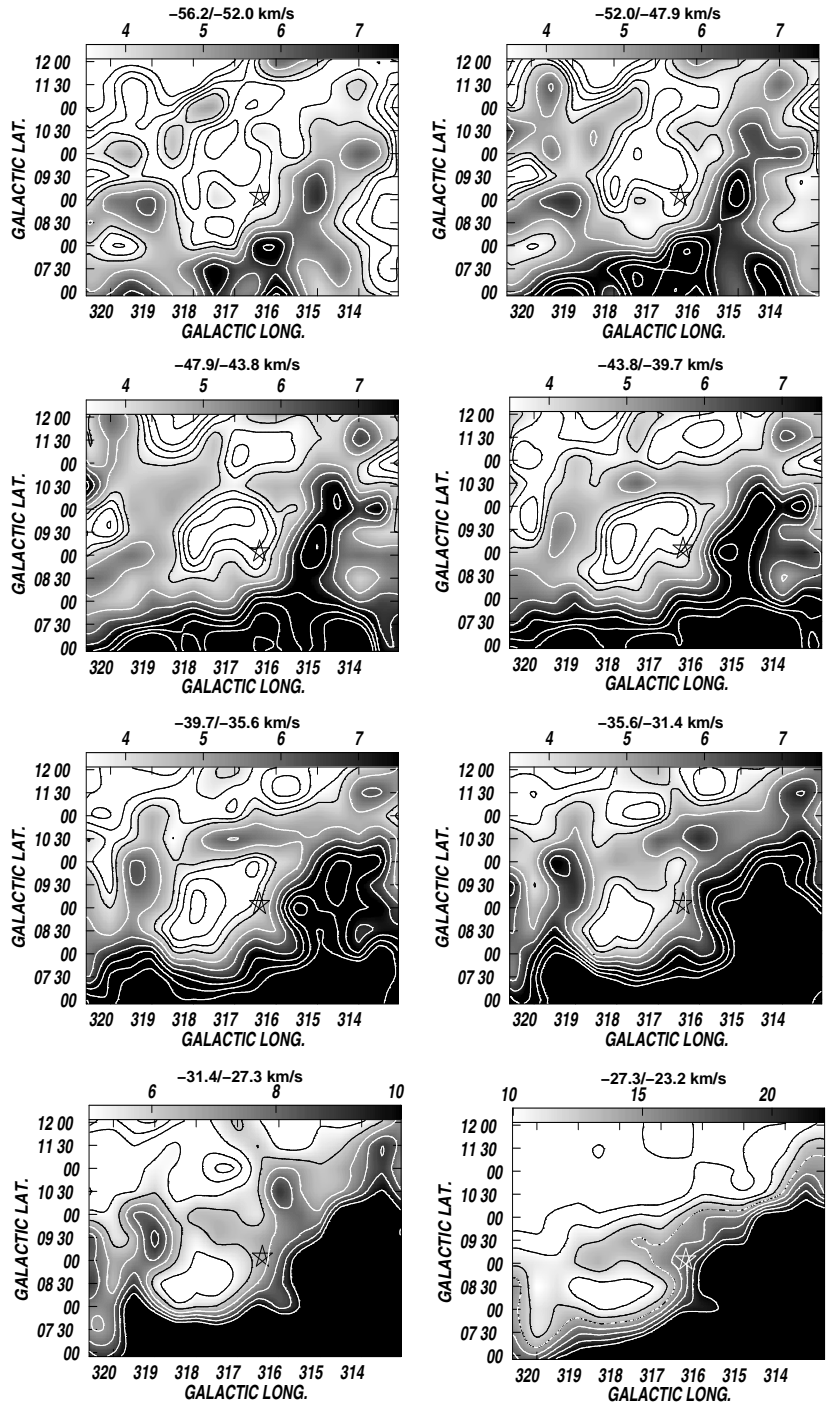}
   \caption{\hi\ emission distribution towards HD\,124979, as in Figure 2. For the seven first maps, the contour lines are from 2 to 4 in steps of 0.5 K and for 5 to 10 in steps of 1 K, and for the last map, from 8 to 22 in steps of 2 K.
   }
    \end{figure}
        
 \begin{figure}
 \centering
          \includegraphics[width=.75\textwidth]{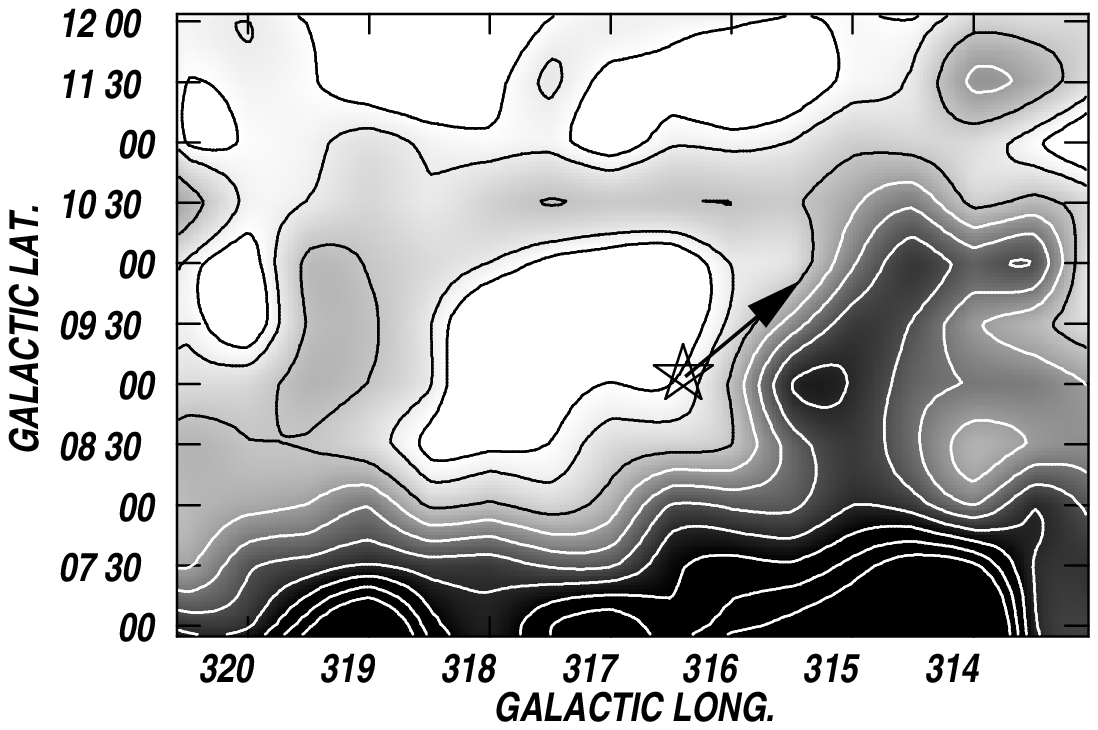}\\
          \includegraphics[width=.75\textwidth]{HD124979-IRIS100.EPS}\\
            \includegraphics[width=.75\textwidth]{HD124979-HALFA.EPS}\\
                \caption{{\it Upper panel.} Probable \hi\ structure related to HD\,124979. The 
image shows the \hi\ column density distribution within the velocity interval
--47.9, --34.6 \kms. The grayscale is from 3.5 to 10 K, and the contour lines 
are from 3.5 K to 12 K in steps of 1 K. {\it Central panel.} Overlay of the far IR emission distribution at 100 $\mu$m 
(grayscale from 21 to 35 MJy ster$^{-1}$) and the \hi\ column density distribution 
{\it (contour lines).}
{\it Bottom panel.} Overlay of the H-Alpha Composite image (grayscale) from 10 to 14 R and the \hi\ column density distribution {\it (contour lines)}.
}
    \end{figure}


The cavity and surrounding shell around this star is shown in Fig. 3 (upper panel). The image was obtained by integrating the \hi\ emission within 
the velocity interval from --7.7 to +14.0 \kms. The feature is elongated 
in the direction of the stellar proper motion, which is marked in the 
figure with a black arrow. 
The central panel displays an overlay of the IRIS image at 100 $\mu$m 
(grayscale) and the \hi\ image (contour lines). The \hi\ shell can partially be detected in the far infrared. 
The presence of a strong IR filament at b \ap\ --25\deg is 
compatible with the existence of a thin \hi\ wall in this region. 
An IR emission band is also projected onto the cavity. If related to the 
structure, 
this emission might be linked to the approaching or receding caps. 

The bottom panel shows a superposition of the same \hi\ contour lines and 
the H-Alpha Composite image. Optical emission is present inside the cavity 
and appears bordered by the neutral envelope towards lower galactic 
longitudes and higher negative galactic latitudes.

Neither catalogued \hii\ regions nor supernova remnants were found over the
regions under study. The image
at 408 MHz (not shown here) displays a ring-like structure centered at {\it (l,b)}
= (235\fdg 0,--28\fdg 0), with inner and outer semiaxes of 
2\fdg 1$\times$0\fdg 7 and 4\fdg 2$\times$1\fdg 9, respectively.
The radio continuum feature is clearly larger than the \hi\ cavity, and most 
of it is projected onto the neutral envelope, contrary to what is 
expected for an interstellar bubble or \hii\ region. Thus, the lack of 
coincidence between the radio continuum feature and the \hi\ cavity casts
doubts on its relation to the neutral shell. 

The systemic velocity of the structure, defined as the velocity at which 
the cavity has its larger dimensions and is better defined, is +8 \kms. 
According to the analytical fit by \citet{BB93}, the low systemic 
velocity derived for the \hi\ structure indicates a kinematical distance 
of about 1.0 kpc, in agreement with the stellar distance listed in Table 1. 
The location of the star with cavity, together with the agreement 
between the kinematical and spectrophotometric distances strongly suggest 
that the \hi\ structure is related to HD\,38666.
We have adopted a distance $d$ = 0.5$\pm$0.1 kpc for this structure.

At this distance, the stellar tangential velocity is quite large, 
$\rm{v_t} \approx$ 57$\pm$15 \kms, and the massive star could have created 
a bow-shock structure. The presence of a bow-shock like object related
to this star was investigated by \citet{vBMc88}, who were not
succesful in finding such structure in the IRAS images. 
 
\subsection{The ISM around HD\,124979}

The analysis of the \hi\ emission distribution maps towards HD\,124979 allows identification of a cavity and shell probably associated with the star.
The \hi\ minimum is detected within the velocity interval from --52.0 to --23.2 \kms, decreasing in angular size for velocities $\rm{v}$ $>$ --31.4 \kms. At the stellar position, the contour lines are distorted in the velocity range --52.0 to --39.7 \kms.

The \hi\ minimum is better defined from --52.0 to --31.4 \kms, and thus the systemic velocity of the structure can be 
derived as the central value of the mentioned interval, $\rm{v_{sys}} \approx$ --42$\pm$10 \kms. Taking into account a velocity dispersion of 6 \kms, 
the systemic velocity corresponds to a kinematical distance of 3.0$\pm$1.0 kpc, which is compatible with the spectrophotometric distance.
We have adopted $d$ = 3.5$\pm$1.0 kpc as the distance to the \hi\ structure.

Figure 5 (upper panel) shows the \hi\ brightness temperature distribution  towards 
HD\,124979, averaged in the velocity interval from
--47.9 to --35.6 \kms, for which the \hi\ cavity is better defined. 
In order to delineate the cavity we have considered the brightness temperature contour corresponding to 3.5 K.  The  centroid 
of the structure is {\it (l,b)} = (317\fdg 2,+9\fdg 3). Both the cavity and the almost complete shell are elongated in the direction of the tangential motion of the star (marked with a black arrow).

    \begin{figure}
   \centering
          \includegraphics[width=.95\textwidth]{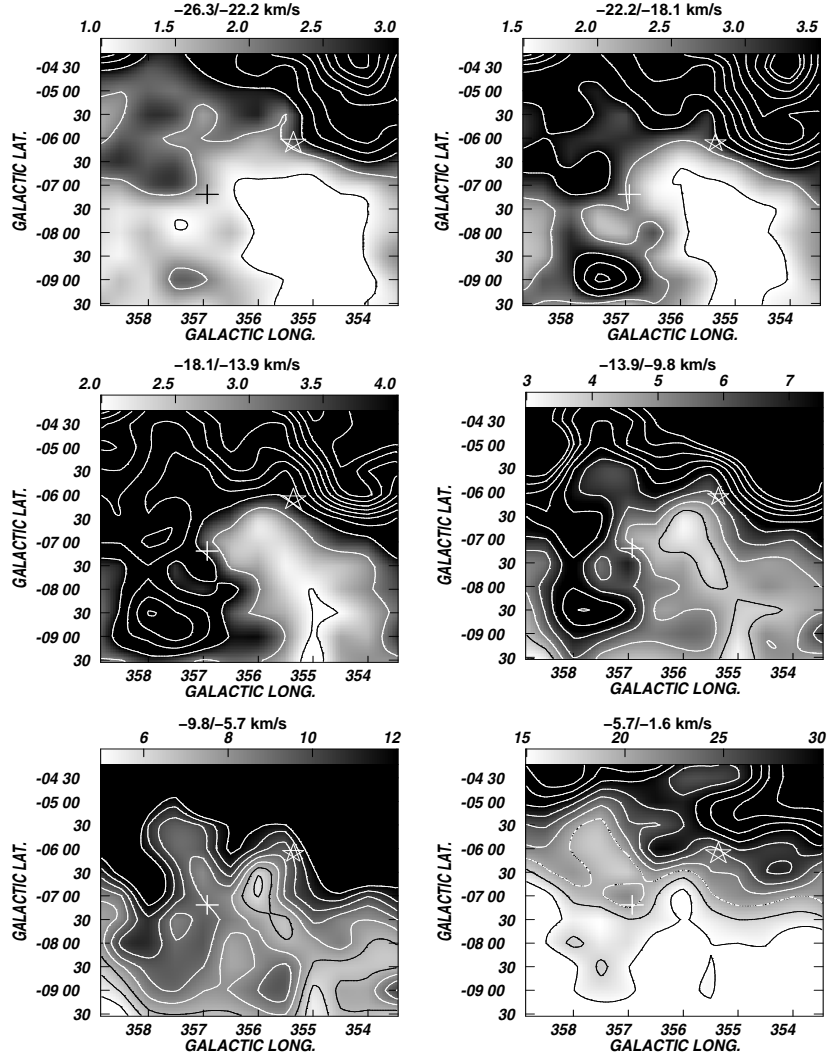}
   \caption{\hi\ emission distribution towards HD\,163758, as in Figure 2. For the maps corresponding to the interval  --26.3 to --5.7 \kms, the 
contour lines are from the minimum value of the grayscale to 12 K, in steps 
of 1 K.
For the image showing the interval --5.7 to --1.6 \kms, the contour lines 
are from the minimum value of the grayscale to 40 K, in steps of 2.5 K. HD\,163658 and WR\,109 are indicated by the star and the cross,
respectively.  }
    \end{figure}

   \begin{figure}
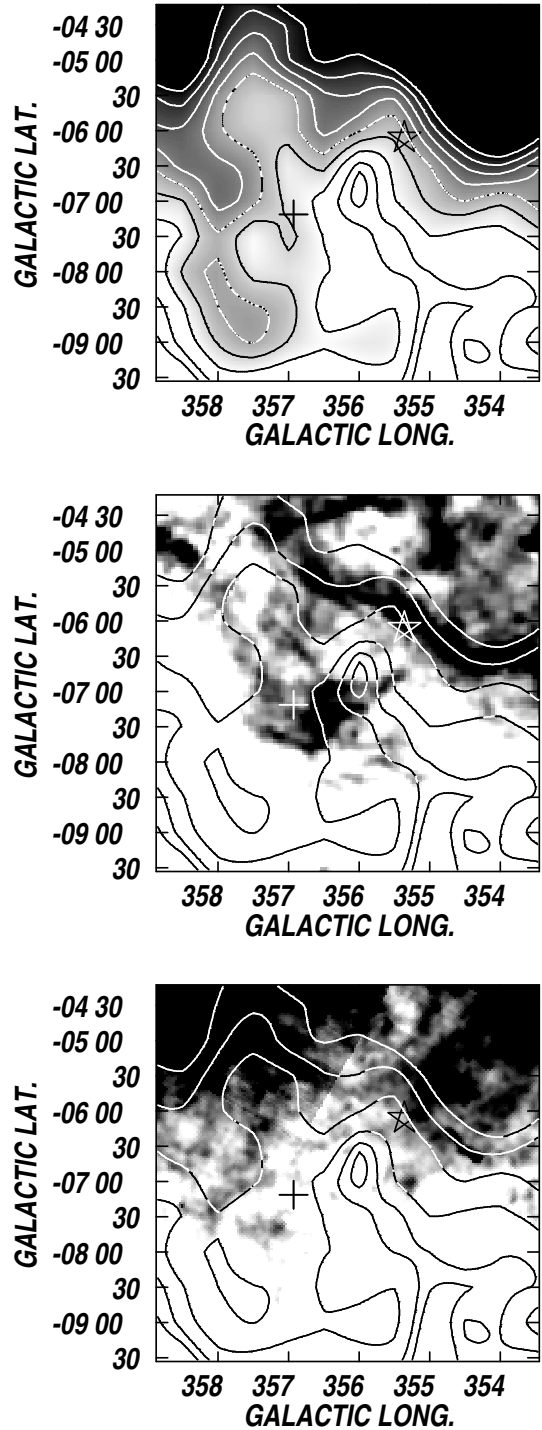

   \centering
          \includegraphics[width=.65\textwidth]{HD163758-MAPS2.EPS}\\
          \includegraphics[width=.65\textwidth]{HD163758-HALPHA-HI.EPS}\\
          \includegraphics[width=.65\textwidth]{HD163758-IRIS-100.EPS}\\
   \caption{{\it Upper panel:} \hi\ emission towards HD\,163758 integrated
within the velocity range --22.2 to --5.7 \kms. The grayscale is from 6 to 
12 K. The contour lines are 4.5 and from 5 to 12 K, in steps of 1 K. {\it Central panel:} Overlay of the H-Alpha Composite image (grayscale from 32 to 45 R) 
and the \hi\ emission (contour lines, 4.5, 5 and from 6 to 12 K, in steps of 2 K). {\it Bottom panel:} Overlay of the 
IR emission at 100 $\mu$m (grayscale from 50 to 80 MJy ster$^{-1}$) and the
\hi\ emission contour lines of central panel.}
    \end{figure}

The 100 $\mu$m IRIS image shows far IR emission bordering the region
towards the galactic plane, revealing the presence of dust 
associated with the bright portions of the surrounding shell (Fig. 5, central panel).
The bright IR areas detected inside the cavity, which extend well beyond the cavity, are probably unrelated to the \hi\ structure.
The proper motion of the star corresponds to a tangential velocity of
$\rm{v_t} \approx$ 200$\pm$30 \kms. A careful inspection of the IRIS
images does not allow to detect a bow-shock like structure at IR wavelengths.

Neither catalogued  \hii\ regions nor supernova remnants are linked to 
the structure. The bottom panel of Fig. 5 indicates that no significant optical 
emission is detected towards this region. An inspection of the image at 408 MHz
do not show identifiable
radio emission probably related to the \hi\ structure.  

No additional OB stars at a compatible distance are 
projected onto the \hi\ structure. 

   \begin{figure}
   \centering
	  \includegraphics[width=.95\textwidth]{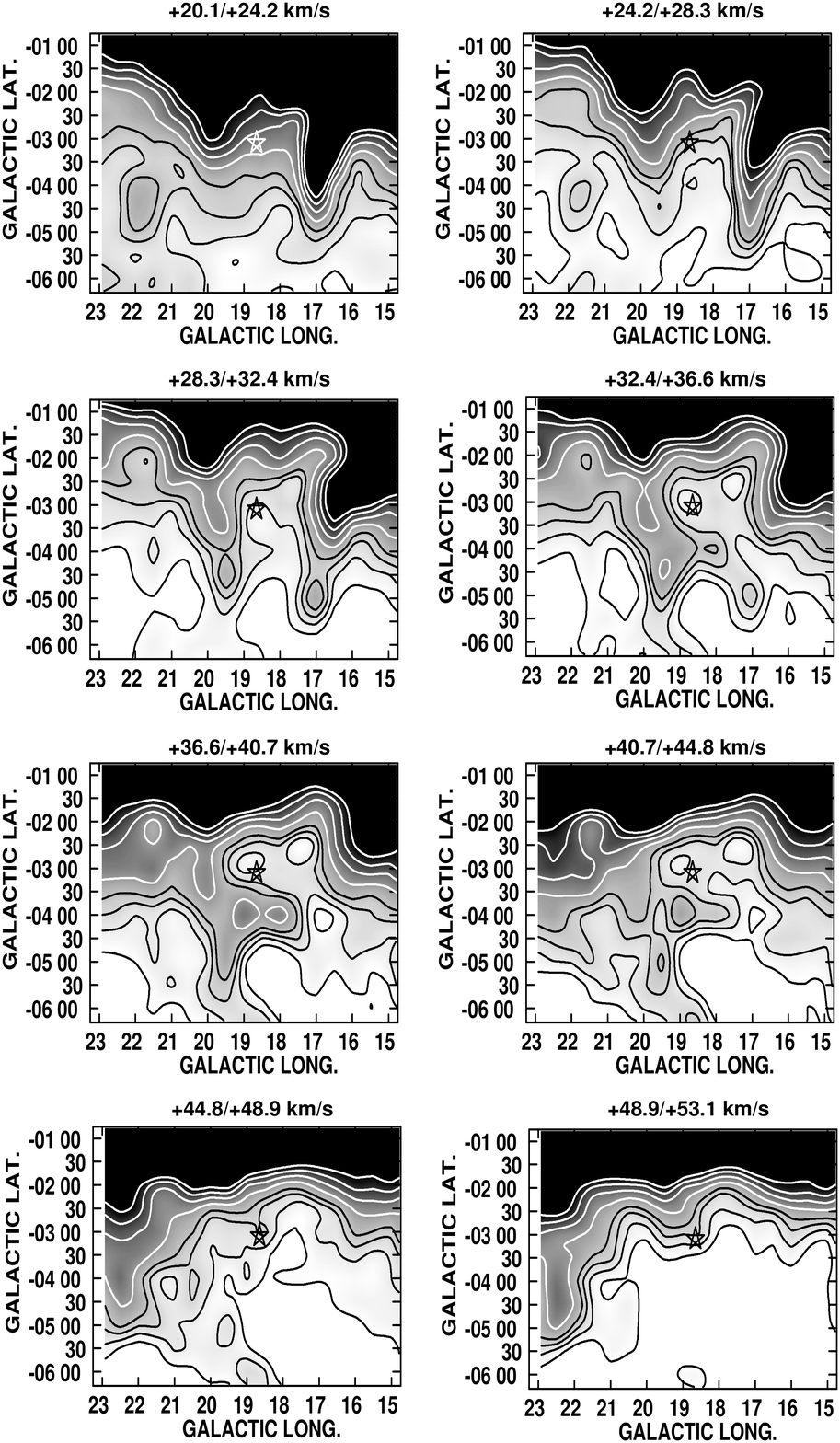}
   \caption{\hi\ emission distribution towards HD\,171589, as in Figure 2. The grayscale is from 5 to 30 K. The contour lines are from 5 to 11 K, in steps of 2 K, and from 15 to 30 K, in steps of 5 K.
   }
    \end{figure}    
   \begin{figure}
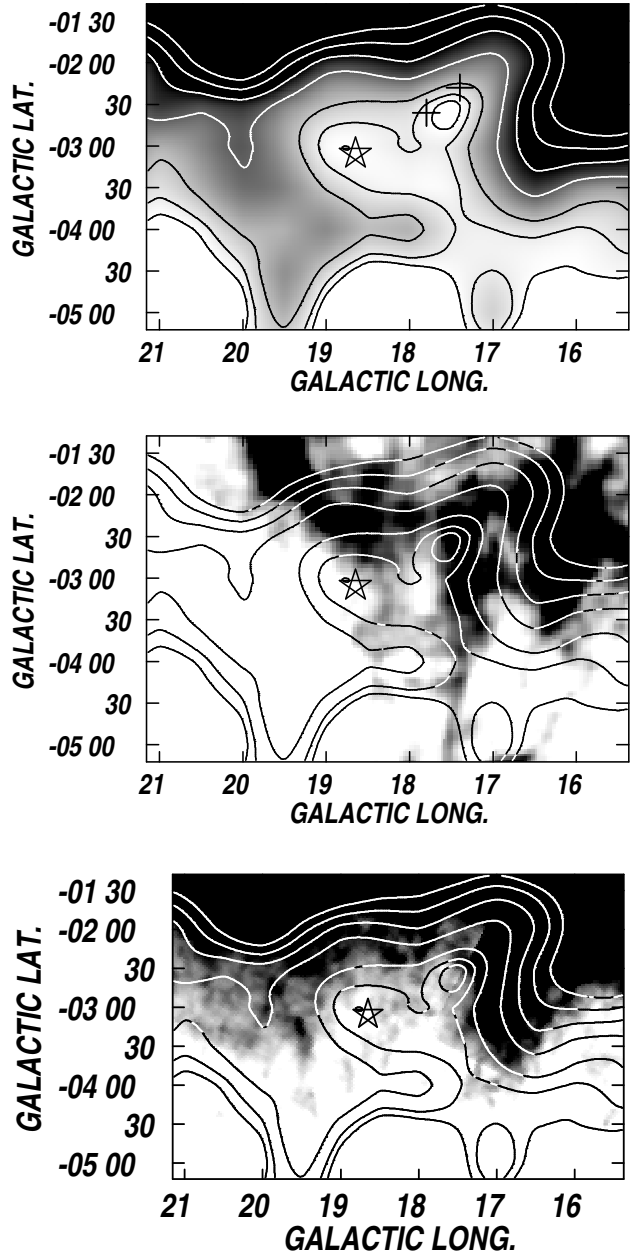

   \centering
          \includegraphics[width=.75\textwidth]{HD171589-MAPS2.EPS}\\
          \includegraphics[width=.75\textwidth]{HD171589-HALPHA-HI.EPS}\\
          \includegraphics[width=.75\textwidth]{HD171589-IRIS-100-3.EPS}\\
   \caption{{\it Upper panel:} \hi\ column density distribution towards 
HD\,171589, within the  velocity range +26.3 to +44.8 \kms. The grayscale 
is from 7 to 20 K. Contour lines are 7, 8, and 10 K, and from 15 to 30 K,
in steps of 5 K. HD 171589 is indicated by a five pointed star, and the 
SNR G17.4--2.3 and SNR G17.8--2.6 are marked by crosses. {\it Central panel:} Overlay of 
the H-Alpha Composite image (grayscale from 35 to 70 R) and the \hi\ 
column density distribution shown in the upper panel (contour lines). 
{\it Bottom panel:} Overlay of the IR emission distribution at 100 $\mu$m 
(grayscale  from 70 to 130 MJy ster$^{-1}$) and the \hi\ column density 
distribution shown in the upper panel (contour lines).   }
    \end{figure}    

\subsection{The ISM around HD\,163758}

Figure 6 displays \hi\ brightness temperature images corresponding to the 
velocity interval from --26.3 to --1.6 \kms\ in the vicinity of HD\,163758.
An inspection of the images shows the presence of a low emission region 
at {\it b} $\leq$ --5\deg.

The \hi\ cavity centered at {\it (l,b)} = (356\fdg 0, --6\fdg 5) can be appreciated at velocities in the range $\rm{v}$ from --26.3 to --5.7 \kms. HD\,163758, which is indicated by the star symbol, appears projected near the higher density border of the cavity, indicated by the piled-up of the contour lines.

The upper panel of Fig. 7 displays the mean \hi\ brightness temperature in 
the range (--22.2,--5.7) \kms, where the cavity and shell centered at {\it(l,b)} = (356\fdg 0, --6\fdg 5) are clearly identified in
contours and grayscale. 
The central and bottom panels show overlays of the \hi\ emission distribution and the H$\alpha$ and 100 $\mu$m IRIS emissions, respectively. The H$\alpha$ image reveals a shell-like structure 
of about 1\fdg 5 in diameter centered at {\it (l,b)} = (356\fdg 0, --6\fdg 5), 
with the O-type star projected onto one of its borders. The IRIS image shows the lack of IR emission in the region of the \hi\ cavity. Bearing in mind the angular resolution of the \hi\ data, the correlation of 
the optical feature with the small \hi\ structure is good, suggesting 
that the \hi\ feature is the neutral gas counterpart of the optical shell. 
The fact that the star appears projected onto the highest density border 
of the \hi\ structure reinforces the association between the star and the cavity.

A search for other massive stars projected onto the region shows that 
the Wolf-Rayet star WR\,109 (= V617 Sgr, WN5h+?, \citealp{vdH01}) is 
placed about 2\deg\ far away from the position of the O star.  According to 
\citet{vdH01}, WR\,109 is located at a photometric distance  of 
34 kpc. The faraway estimated distance is the result of the low optical 
absorption derived for the star, along with its low apparent magnitude. 
This WR star, marked by a cross symbol in Figs. 6 and 7, is also near the 
border of the cavity. Its position well outside the optical shell suggests 
that the WR star is unconnected to the H$\alpha$ shell.

\begin{table}[!t]\centering
\small
  \newcommand{\DS}{\hspace{6\tabcolsep}}
  \begin{changemargin}{-2cm}{-2cm}
    \caption{Physical parameters of the \hi\ structures around the O stars}             
  \setlength{\tabnotewidth}{0.95\columnwidth}
    \setlength{\tabcolsep}{1.2\tabcolsep} \tablecols{5}
    \begin{tabular}{l @{\DS} cccc l cccc}
\toprule            
      &HD\,38666 &  HD\,124979 & HD\,163758 & HD\,171589  \\
\midrule
{\it (l,b)} center& (235\fdg 5,--27\fdg 5)& (317\fdg 2,+9\fdg 3)& (356\fdg 0,--6\fdg 5) & (18\fdg 7,--3\fdg 0)\\
Velocity interval $\rm{v}_1,\rm{v}_2$ (\kms) & --7,+17 & --53,--25 & --22,--5  & +26,+45    \\
Systemic velocity $\rm{v}_{sys}$ (\kms) & +8$\pm$1 &  --42$\pm$3 &  --10$\pm$1 & +34$\pm$1  \\
Expansion velocity $\rm{v}_{exp}$ (\kms) & 11 &  15 &  10  &   11  \\
Kinematical distance (kpc) & 1.0  & 3.0$\pm$1.0 & 3.3$\pm$1.0 & 3.0$\pm$0.5    \\
Adopted distance (kpc) & 0.5$\pm$0.1 &  3.5$\pm$1.0 & 3.5$\pm$1.0 &3.0$\pm$0.6 \\
Radius of the \hi\ cavity $R_{cav}$ & 1\fdg 6 &   1\fdg 25 &  0\fdg 6  & 0\fdg 6  \\
Radius of the \hi\ structure $R_{s}$ & 3\fdg 0 &   1\fdg 9 &  1\fdg 2 & 1\fdg 5  \\
Radius of the \hi\ structure $R_{s}$ (pc) & 26$\pm$5 &  116$\pm$33 &  70$\pm$20 & 52$\pm$9 \\
Mass in the shell (\msun) & 600$\pm$240 & 58800$\pm$33600 & 4300$\pm$2450 &  6900$\pm$2300  \\
Mass deficiency (\msun) & 80$\pm$32 & 6000$\pm$3400 &  1100$\pm$630 &  350$\pm$120 \\
Swept-up mass $M_{s}$ (\msun)  &  340$\pm$140  & 32400$\pm$18500 & 2700$\pm$1540 & 3600$\pm$1200 \\ 
$n_e$ (\cmtres) (f=1.0) & \nodata  & 0.5 & \nodata & 0.9 \\
$M_i$ (\msun) (f=1.0) & \nodata & 24000 & \nodata & 2900 \\
$n_e'$ (\cmtres) (f=0.3) & \nodata & 0.9 & \nodata & 1.8 \\
$M_i'$ (\msun) (f=0.3) & \nodata & 13200 & \nodata & 1600 \\
Kinetic energy (10$^{48}$ erg)  & 0.4  &  73-103 & 2.7  & 4.4-6.3 \\
Dynamical age (10$^{6}$ yr)  & 1.2 & 4.3 & 3.5 & 3.8 \\
Ambient density $n_o$ (cm$^{-3}$)  &   0.2 & 0.2-0.3  & 0.08  & 0.25-0.35  \\
\bottomrule      
\end{tabular}
   \end{changemargin}
\end{table}

The systemic velocity of the \hi\ structure is --10$\pm$1 \kms. The 
analytical fit to the circular galactic rotation model predicts that 
material at velocities of about --10 \kms\ is placed at near and far kinematical 
distances of 3.3$\pm$1.0 and 13-14 kpc, respectively. The near kinematical distance agrees with
the spectrophotometric distance to HD\,163758 (see Table 1). Note that 
kinematical distances have large uncertainties in this section of the Galaxy. 
We adopt 3.5$\pm$1.0 kpc as the distance to the \hi\ feature.

According to SIMBAD, no \hii\ regions or SNR have been identified in this
area. No additional OB stars were found projected onto the \hi\ 
structure.  
The analysis of the CO images by \citet{Dame01} did not show detectable 
molecular emission in this region.

\subsection{The ISM around HD\,171589}

Figure 8 displays the \hi\ emission distribution within the velocity 
interval from +20.1 to +57.2 \kms\ in a large area around the massive star. 
Two cavities are clearly detected for velocities in the range +26 to +45 
\kms, centered at {\it (l,b)} = (18\fdg 7,--3\fdg 0) and 
(17\fdg 5,--2\fdg 7). HD\,171589 appears projected onto the  cavity at larger
galactic longitudes. 

The upper panel of Fig. 9 displays the \hi\ column density distribution 
within the velocity range from +26.3 to +44.8 \kms. Both cavities are separated 
in this image. The O-type star is projected onto the \hi\ hole at 
{\it (l,b)} = (18\fdg 7,--3\fdg 0), partially surrounded by an \hi\ shell.

The central and bottom panels of Fig. 9 show the superposition of the \hi\ 
column density image (in contours) and the H$\alpha$ and IRIS images, respectively. 
The central panel reveals a region without optical emission coincident with 
the cavity at {\it (l,b)} = (18\fdg 7,--3\fdg 0). The bottom panel shows that both \hi\ holes are partially outlined by bright IR emission at {\it b} $>$ --3\fdg 5. The distribution of 
the \hi, optical, and IR emissions is compatible with the presence of an 
interstellar bubble driven by the massive star. 

The systemic velocity of the structure at {\it (l,b)} = (18\fdg 7,--3\fdg 0)
is +36$\pm$2 \kms, corresponding to near and far kinematical distances 
of 3.2$\pm$0.5 and 13 kpc, respectively. The near kinematical distance agrees with
the spectrophotometric distance derived for the star. Consequently, we 
adopt $d$ = 3.0$\pm$0.5 kpc.

Six supernova remnants were detected in this area at galactic latitudes 
$b \geq$ --2.8 \citep{Green04}. Two of them (\hbox{G17.4--2.3} and 
\hbox{G17.8--2.6}) appear projected onto the cavity centered at {\it (l,b)} 
= (17\fdg 5,--2\fdg 7). Based on the $\Sigma$-D relation, \citet{Gu03} derived distances of about 6.3 kpc to 
the SNRs. As discussed by many authors (e.g. Green 2005) distances derived from the $\Sigma$-D relation are hardly reliable. Both SNRs are detected at 
35 MHz \citep{Dwarakanath90}. An inspection of the image at
4850 MHz shows weak emission also probably related to these remnants. Two different facts can be proposed as the origin of the \hi\ cavity. 
On one hand, the \hi\ cavity and shell can be associated with the SNRs, as 
was found in many SNRs  (e.g. \citealp{Paron2006,Reynolds2008}).  
In this case its systemic velocity of +36 \kms\ suggests near and far 
kinematical distances of 3.2 and 13 kpc, respectively. 
On the contrary, if the \hi\ hole originates in absorption due to the 
radio continuum sources, the distance to the SNRs can be inferred from the 
highest positive velocity at which the hole is detected (about +52 \kms).
This suggests that the SNRs are placed at distances $\geq$ 4.5 
kpc. Further data are necessary to elucidate this question.

CO emission (\citealp{Dame01}) is present mainly at {\it b} $>$ --3\deg. 
The integrated CO emission (not shown here) in a velocity range similar to 
that of the
\hi\ emission does not show molecular material linked to the hole at 
{\it (l,b)} = (17\fdg 5,--2\fdg 7). Weak CO emission present in the 
interval +28.0 to +43.6 \kms\ encircles the border of the cavity related to the SNRs towards {\it b} $>$ --3\deg. 

\section{Discussion}

\subsection{Physical parameters}

The main physical parameters of the neutral gas structures linked to
the O-type stars are summarized in Table 2. 
The {\it (l,b)} centers correspond to the approximate centroid of the 
features. The velocity interval indicates the velocity range where the 
\hi\ structures can be identified, being $\rm{v_{1}}$ and $\rm{v_{2}}$ 
the lowest and highest velocities, respectively, at which the features are detected. 

Following \citet{Cappa08}, the expansion velocities were estimated as $\rm{v_{exp}} = (\rm{v_{2}} -  \rm{v_{1}}$)/2 + 1.3 \kms. The derived values are lower limits since the caps of the expanding shells are not detected in the present cases. The extra 1.3 \kms\ takes into account the fact that the caps may be present just outside the velocity range at which the \hi\ cavity is detected, that is at velocities $\rm{v_{1}}$ - 1.3 \kms\ and $\rm{v_{2}}$ +1.3 \kms.
  
For the four selected stars the adopted distances are compatible with the 
spectrophotometric and kinematical distance estimates. 
Uncertainties in the adopted kinematical distances for HD\,124979,
HD\,163758, and HD\,171589 arise in a velocity
dispersion of $\pm$6 \kms\ adopted for our Galaxy. 

\begin{table}[!t]\centering
\small
  \newcommand{\DS}{\hspace{6\tabcolsep}}
  \begin{changemargin}{0cm}{-2cm}
\caption{Energetics of the structures}             
  \setlength{\tabnotewidth}{0.95\columnwidth}
    \setlength{\tabcolsep}{1.2\tabcolsep} \tablecols{5}
    \begin{tabular}{l @{\DS} cccc l cccc}
\toprule                
       & HD\,38666 &  HD\,124979 & HD\,163758 & HD\,171589  \\
\midrule
T$_{eff}^{\mathrm{a}}$    & 34600  & 37100  & 40200  & 40000  \\
$ log \ L$ (\lsun)        & 5.0  & 5.6    & 6.03   & 5.7 \\
$M$ (\msun)               & 23  & 40     & 70     & 47  \\
$\dot{M}$ (10$^{-6}$ \mdot) & 0.00032$^{\mathrm{b}}$-0.005$^{\mathrm{c}}$ & 1.8 $^{\mathrm{e}}$ & 8.3$^{\mathrm{e}}$ & 2.7$^{\mathrm{e}}$ \\
 & 0.02$^{\mathrm{d}}$-0.35$^{\mathrm{e}}$  & & & \\
V$_{\infty}$ (\kms)         & 1200$^{\mathrm{b}}$  & 2100$^{\mathrm{f}}$ & 2400$^{\mathrm{f}}$ & 2500$^{\mathrm{f}}$ \\
L$_{w}$ (10$^{36}$ erg s$^{-1}$) &  0.0002-0.16 & 2.5 & 15.0 & 5.4 \\
Dynamical age (10$^6$ yr)          &    1.2   & 4.3 & 3.5  & 3.8  \\
E$_w$ (10$^{48}$ erg)            &  0.008-6.0 & 350 & 1660 & 660 \\
$\epsilon$  & 50-0.07 & 0.2-0.3 & 0.002  & 0.007-0.01 \\
Stellar lifetime$^{\mathrm{g}}$ (10$^6$ yr) & 7.0 & 4.5 & 3.0 & 4.0 \\
\bottomrule                                 
    \tabnotetext{a}{\citet{Va96}}
       \tabnotetext{b}{\citet{Martins05}}
    \tabnotetext{c}{\citet{CG91}}
    \tabnotetext{d}{\citet{Howarth1989}}
     \tabnotetext{e} {Estimated from \citet{Vink00}}
        \tabnotetext{f}{Adopted from \citet{Pr90}}
        \tabnotetext{g}{\citet{Schaller92}}
\end{tabular}
  \end{changemargin}
\end{table}


The radius of each cavity corresponds to the geometric mean of 
the major and minor semiaxes, while the radii of the \hi\ shells 
($R_s$) were evaluated from
the position of the maxima in the envelopes.
Errors in radii come from the distance uncertainty.
 
Neutral atomic masses in the shells and mass deficiencies in the cavities are also listed in Table 2. The structures surrounding
HD\,124979 and HD\,171589 are easier to define over the halves further 
away from the galactic plane. In each case, the other half is contaminated
with diffuse emission from the galactic plane. Thus, for these two
structures, the neutral mass in the shells were derived as twice the mass 
associated with the better defined half. The sewpt-up neutral masses were obtained as mean values between the neutral mass deficiency in the voids and the neutral mass in the envelopes. This procedure allows to remove a first order contribution of the background emission, since both mass determinations are probably contaminated with neutral gas unrelated to the structures. Values listed in Table 2 include a typical interstellar He abundance of 10\%.

In the previous sections we have stated that no radio continuum emission 
associated with the \hi\ structures was detected. Assuming that the massive 
stars have ionized the surrounding 
gas through their strong UV photon flux, and that this ionized material has been swept-up and is present in the inner borders of the neutral envelopes, we can derive an upper limit for the electron density n$_{e}$ and the ionized mass M$_{i}$ from the rms flux density at 4.85 GHz (7.7 mJy beam$^{-1}$). 
Adopting $R_{ou}$ = 1.0 $R_{cav}$ and $R_{in}$ = 0.9 $R_{cav}$ as the outer and inner radii of 
the ionized regions, we estimated upper limits for the flux densities S$_{4.85 GHz}$ =  3.7 and 1.0 Jy for the ionized
regions around HD\,124979 and HD\,171589, respectively. Estimates of the physical parameters of the \hii\ regions towards these stars
can be obtained using the expressions by \citet{MH67} for 
spherical ionized regions of constant density. 
Adopting an electron temperature of 8$\times$10$^{3}$ K and a volume filling factor $f$ = 1.0, we derived the electron density $n_{e}$ and the associated ionized mass $M_{i}$. A different electron density and ionized mass can be estimated by considering an alternative filling factor. For an ionized shell with outer and inner radii of $R_{ou}$ and $R_{in}$, in which the plasma covers an area A equal to 50\% of the surface of the shell, the filling factor can be derived as $f$ \ = \ A $\times$\ ($R_{ou}^{3}$ - $R_{in}^{3}$)/$R_{ou}^{3}$ = 0.3. 
This value was used to derive $n_{e}'$ and $M_{i}'$. The derived values for $n_{e}'$ are consistent with the high $z$-distances of the stars.
Note that electron densities and ionized masses are upper limits.
Unfortunately, the lack of radio data at frequencies higher than 1 GHz precludes from deriving upper limits for the regions of HD\,38666 and HD\,163758.

The kinetic energy $E_\mathrm{k}$ = $M_\mathrm{b}\rm{v_{exp}}^2/2$ 
was estimated from the expansion velocities listed in Table 2 and the 
neutral and ionized masses in the structures. The range in kinetic 
energies corresponds to the fact that we have taken into account the 
neutral atomic mass only, and the neutral atomic and ionized masses.

Dynamical ages were estimated as $t_\mathrm{d}$ = 
0.55$\times$10$^{6}$ R$_{s}$/\rm${v_{exp}}$ yr \citep{McCray83}, where 
R$_{s}$ is the radius of the bubble (pc), $\rm{v_{exp}}$ is the expansion 
velocity (\kms), and the constant represents a mean value between the energy 
and the momentum conserving cases.

Finally, ambient densities $n_o$ were derived by uniformly distributing
the associated mass within a sphere of radius R$_{s}$. For the regions of HD 124979 and HD 171589 two values are listed in Table 2. The first one was obtained by distributing the swept-up neutral mass ($M_{s}$) while the second one was derived by distributing the neutral and ionized masses ($M_{s}$ + $M_{i}'$). The major source of error in radii and masses is the distance uncertainty.  
The low ambient densities are consistent with the structures being far
from the galactic plane. 

\subsection{Energetics and origin of the structures}

We will analyze here whether the 
massive stars can provide the energy to create the cavities and shells 
found in the previous sections through their stellar winds and UV photon
fluxes.
To test the former  possibility, we will estimate the mechanical energy 
$E_\mathrm{w}$ released into the ISM for the massive stars and compare it 
with the kinetic energy  $E_\mathrm{k}$ of the structures.

Table 3 summarizes relevant parameters useful to evaluate the energetics 
of the structures. The values of effective temperature $T_{eff}$, stellar 
luminosity $L$, and stellar mass listed in the first three rows correspond to the spectral classification and luminosity class of the O-type stars listed in Table 1. These data were used 
to derive the mass loss rate $\dot{M}$ following the recipe by \citet{Vink00}.  
Based on the mass loss rates and terminal velocities, we estimated the 
mechanical wind luminosity for each star as
$L_\mathrm{w} = \dot MV_\mathrm{w}^2/2$.

The stellar wind mechanical energy $E_\mathrm{w} (= L_\mathrm{w}t)$ released 
by the massive stars were derived from the  dynamical ages of the bubbles. The resulting values are included in Table 3.

The ratio $\epsilon$ between the kinetic energy $E_\mathrm{k}$ 
and the mechanical energy $E_\mathrm{w}$ provided by  the massive stars
during the dynamical age of the structures is also included in Table 3. 
Evolutionary models of stellar wind bubbles predict that $\epsilon$ = 0.2 
for the energy conserving case and $\epsilon \leq$ 0.1 for the
momentum conserving case (see \citealp{McCray83}).
Although the uncertainty in this value is large
(at least 70\% adopting a 30\% error in the distance), it is clear
that HD\,163758 and HD\,171589 are capable of blowing the observed
estructures. The result for HD\,124979 is still consistent with an
interstellar bubble interpretation. 

The obtained dynamical ages for HD 124979, HD 163758, and HD 171589 are compatible with the ages derived from evolutionary tracks for stars with solar abundances (see \citealp{Schaller92}).

The case for HD\,38666 is more complex. The large uncertainty in the  
estimated mechanical energy precludes for giving a clear conclusion about 
the origin of this structure. The derived dynamical age is lower than the age estimate obtained from \citet{Schaller92}. However, \citet{Martins05} estimate an age $<$ (2-4)$\times$10$^6$ yr, closer to the dynamical age. 
On the other hand, \citet{Hoogerwerf01} proposed a binary-binary collision between three stars of Trapezium cluster: AE Aur, HD 38666, and $\iota$ Ori , becoming HD 38666 and AE Aur runaway stars. The dynamical ejection scenario took place 2.5$\times$10$^6$ yr ago. We note that a search for other catalogued massive stars at a distance compatible with that of the structure gave negative results.
A different analysis can be done bearing in mind the observed proper
motions.
HD\,38666 and HD\,124979 have large spatial velocities, about 57$\pm$11 \kms\
for HD\,38666, and 200$\pm$30 \kms\ for HD\,124979. Considering
these velocities, which include the uncertainty in the adopted distance, 
it took HD\,38666 about (0.4-0.9)$\times$10$^6$ yr
to cross the \hi\ structure. As regards HD\,124979, the time necessary
to cross the structure was about (0.6-1.0)$\times$10$^6$ yr. 
Adopting 0.7$\times$10$^6$ yr for HD\,38666 and 0.8$\times$10$^6$ yr for
HD\,124979, the mechanical energy turns out to be 
(0.004-3.5)$\times$10$^{48}$ erg for HD\,38666 and 
65$\times$10$^{48}$ erg for HD\,124979. The large spatial velocity of the runaway star HD\,124979 shortenes the crossing time through the structure, which makes their origin to remain unclear. The resulting $\epsilon$-values
indicate that other energy sources are necessary to create both
structures through the stellar wind mechanism. Additional studies are necessary to identify the agents.

No bow-shock like features were found related to HD\,38666 and HD\,124979.
As shown by \citet{Rag97} and \citet{HK01}, bow-shocks
appear associated with only about 30-40\%\ of OB runaway stars. As indicated
by the last authors, large separation from the galactic plane, 
extremely high space velocities and large distances make difficult the
formation and detection of bow-shock structures. Moreover, the physical
conditions of the ambient medium where the star is inmersed play a major
role in determining the existence of bow-shocks. Particularly, these authors find
an anticorrelation between bow-shocks and hot bubbles. Whether the
observed structures originate in the mass flow of the massive stars, the
lack of bow-shocks in these cases is  
consistent with their conclusions.

Taking into account the spectral classification and the luminosity class 
of the associated massive stars and the ambient density where the 
structures are evolving, the estimated radius of the Str\"omgren's spheres 
is in all cases larger than the \hi\ cavities. The smaller size of the cavities can be explained taking into account that a certain amount of UV photons 
are used in dust heating or drain from the patchy envelopes.

\section{Summary}

We have analyzed the interstellar medium in the environs of the O-type stars HD\,38666, HD\,124979, HD\,163758, and HD\,171589. The study of the neutral hydrogen distribution in direction to these stars allowed us to disclose \hi\ structures located at kinematical distances compatible with the stellar distances and probably related to the stars.

Assuming a stellar wind mechanism, the derived dynamical ages for the \hi\ structures related to HD 124979, HD 163758, and HD 171589 are compatible with the lifetimes of the O-type stars in the main sequence, reinforcing the association with the stars.

We have investigated the counterparts of the structures at other frequencies. The surrounding \hi\ shells around HD 38666, HD 124979, and HD 171589 have IR counterparts, revealing the presence of dust associated with the \hi\ gas. The \hi\ shell related to HD 163758 has IR and H$\alpha$ counterparts, with the star projected onto one of its higher density borders.

To investigate the origin of the structures, we have compared the mechanical energy released into the ISM for the massive stars and the kinetic energy of the structures. Our results for the energy conversion efficiency for HD\,163758 and HD\,171589 indicate that the stars are capable of blowing the observed structures. As regards HD\,38666 and HD\,124979, additional energy sources are probably necesary, taking into account the derived $\epsilon$-values and the large tangential velocities. Aditional studies are needed to clarify the origin.

\begin{acknowledgements}
We thank Dr. P. Benaglia for her help in the first stages of this paper. We also thank Dr. J.C. Testori for his collaboration. It is a pleausure to thank the anonymous referee for many comments and suggestions that improve this presentation.
This project was partially financed by the Consejo Nacional de 
Investigaciones Cient\'{\i}ficas y T\'ecnicas (CONICET) of Argentina under 
project PIP 5886/05, Agencia PICT 14018, and UNLP under projects 11/G072. 
\\ 
We acknowledge the use of NASA's SkyView facility (http://skyview.gsfc.nasa.gov) located at NASA Goddard Space Flight Center.
The reduction and analysis of the PMN Survey data was largely the work of 
Mark Griffith and Alan Wright. The FITS maps of the PMN Survey were 
produced by Jim Condon (NRAO) and Niven Tasker.
This research has made use of the SIMBAD database, operated at CDS, Strasbourg, France.
\end{acknowledgements}

\bibliography{mcmartin}

\end{document}